\documentclass{sf2a-conf2017}
\usepackage{graphicx}
\usepackage{hyperref}
\usepackage[]{natbib}  
\usepackage{epstopdf}
\usepackage[utf8]{inputenc}

\def\BibTeX{{\rm B\kern-.05em{\sc i\kern-.025em b}\kern-.08em
    T\kern-.1667em\lower.7ex\hbox{E}\kern-.125emX}}
\bibpunct{(}{)}{;}{a}{}{,}  
\newcommand{\kms}{{\mathrm{km~s^{-1}}}}

\newcommand{\teff}{{\mathrm{Teff}}}
\newcommand{\logg}{{\mathrm{\log g}}}

\begin{document}

\TitreGlobal{SF2A 2017}


\title{HR 7098: a new cool HgMN star ?}

\runningtitle{HR 7098}

\author{R.Monier}\address{LESIA, UMR 8109, Observatoire de Paris Meudon, Place J.Janssen, Meudon, France}
\author{M.Gebran}\address{Department of Physics and Astronomy, Notre Dame University - Louaize, PO Box 72, Zouk Mikael, Lebanon}
\author{F.Royer}\address{GEPI, UMR 8111, Observatoire de Paris Meudon, Place J.Janssen, Meudon, France}
\author{T. K{\i}l{\i}co\u{g}lu}\address{Department of Astronomy and Space Sciences, Faculty of Science, Ankara University, 06100, Ankara, Turkey}


\setcounter{page}{237}


\maketitle


\begin{abstract}
Using one archival high dispersion high quality spectrum of HR 7098 (A0V) obtained with the \'echelle spectrograph SOPHIE at Observatoire de Haute Provence, we show that this star is not a superficially normal A0V star as hitherto thought. The model atmosphere and spectrum synthesis modeling of the spectrum of HR 7098 reveals real departures of its abundances from the solar composition. We report here on our first determinations of the elemental abundances of 35 elements in the atmosphere of HR 7098. Helium and Carbon are underabundant whereas the very heavy elements are overabundant in HR 7098.
\end{abstract}

\begin{keywords}
stars: individual, stars: Chemically Peculiar
\end{keywords}


\section{Introduction}
 HR 7098 currently assigned an A0V spectral type, is one of the 47 northern slowly rotating early-A stars studied by \cite{Royer14}. 
 This star has been little studied: only 47 references can be found in ADS although it is fairly bright (V= 6.63).
 The low projected rotational velocity of HR 7098 can either be due to i) a very low inclination angle ($i \simeq 0$) or ii) a very low equatorial velocity $v_{e}$ . In this second case, the star could develop large over and underabundances and be a new Chemically Peculiar (CP) star.
 We have recently synthesized several lines of 35 elements present in the archival SOPHIE spectrum of HR 7098  using model atmospheres and spectrum synthesis including hyperfine structure of various isotopes when necessary. These synthetic spectra were iteratively adjusted to the archival high resolution high signal-to-noise spectrum of HR 7098 in order to derive the abundances of these elements. This abundance analysis yields underabundances of the light elements He and C, mild overabundances of  the iron-peak elements and large excesses in the very heavy elements (VHE whose atomic number Z is greater than 30). This  definitely shows that HR 7098 should be reclassified as a new CP star. We present here preliminary determinations of the elemental abundances in HR 7098.
  
\section{Observations and reduction}

HR 7098 has been observed at the Observatoire de Haute Provence using the High Resolution (R =75000) mode of SOPHIE in August 2009.
One 20 minutes exposure was secured with a $\frac{S}{N}$ ratio of about 224 at 5000 \AA. 
We did not observe HR 7098 ourselves but fetched the spectrum from the SOPHIE archive.

\section{Model atmospheres and spectrum synthesis }

The effective temperature and surface gravity of HR 7098 were first evaluated using Napiwotzky et al's (1993) UVBYBETA calibration of Stromgren's photometry.
The found effective temperature $\teff$ is 10200 $\pm$ 200 K and the surface gravity $\logg$ is 3.55 $\pm$ 0.25 dex. 

A plane parallel model atmosphere assuming radiative equilibrium, hydrostatic equilibrium and local thermodynamical equilibrium has been first computed using the ATLAS9 code \citep{Kurucz92}, specifically the linux version using the new ODFs maintained by F. Castelli on her website\footnote{http://www.oact.inaf.it/castelli/}. The linelist was built starting from Kurucz's (1992) gfhyperall.dat file  \footnote{http://kurucz.harvard.edu/linelists/} which includes hyperfine splitting levels.
This first linelist was then upgraded using the NIST Atomic Spectra Database 
\footnote{http://physics.nist.gov/cgi-bin/AtData/linesform} and the VALD database operated at Uppsala University \citep{kupka2000}\footnote{http://vald.astro.uu.se/~vald/php/vald.php}.
A grid of synthetic spectra was then computed with a modified version of SYNSPEC49 \citep{Hubeny92,Hubeny95} to model the lines. The synthetic spectrum was then convolved with a gaussian instrumental profile and a parabolic rotation profile using the routine ROTIN3 provided along with SYNSPEC49.
We adopted a projected apparent rotational velocity $v_{e} \sin i =  10.5 $ km.s$^{-1}$ and a radial velocity $v_{rad} = -10.84 $ km.s$^{-1}$ from \cite{Royer14}.

\section{Determination of the microturbulent velocity}

\begin{figure}[h!]
 \centering
 \includegraphics[width=0.5\textwidth]{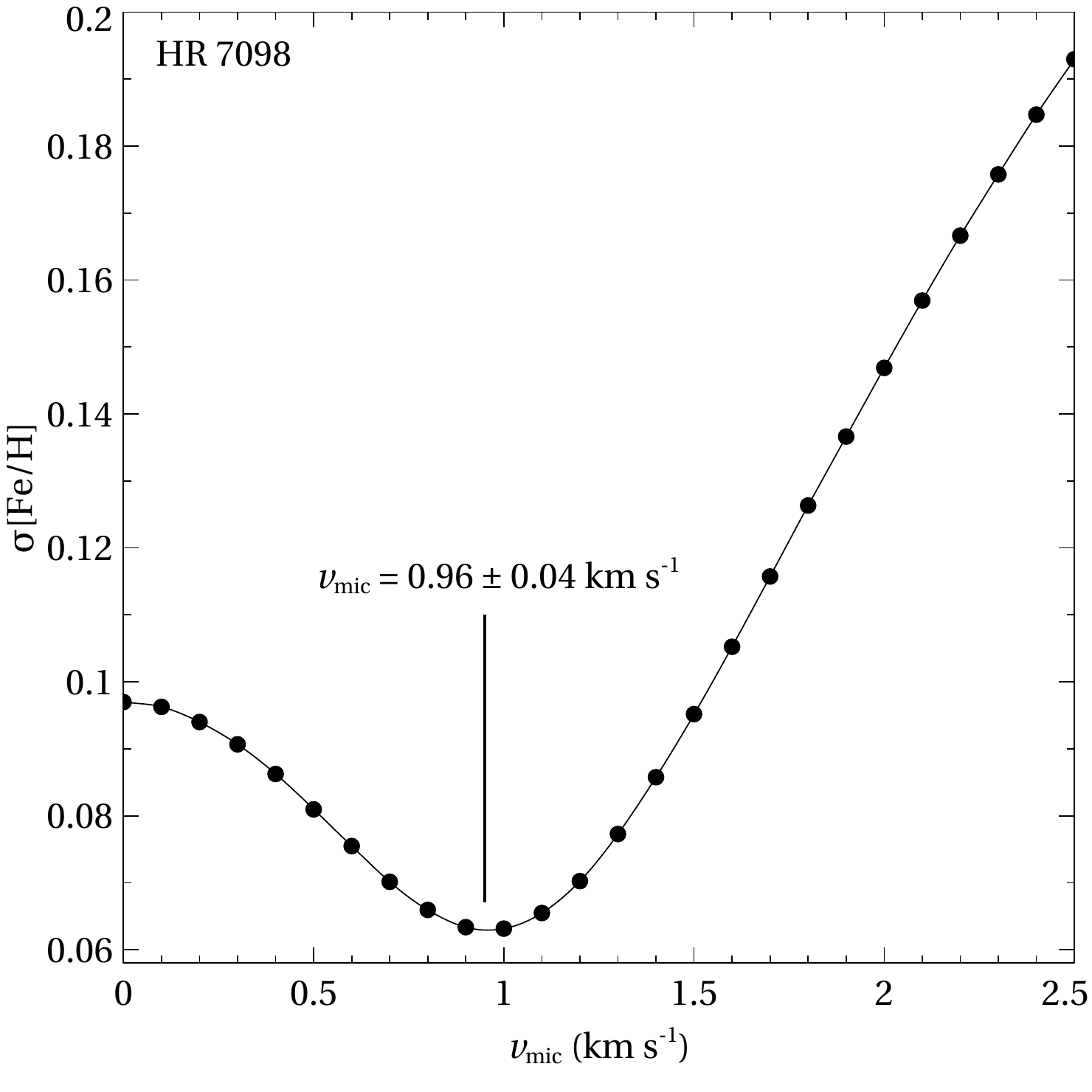}     
  \caption{The derived microturbulent velocity for HR 7098}
  \label{fig1}
\end{figure}

In order to derive the microturbulent velocity of HR 7098, we have derived the iron abundance
[Fe/H] by using 36 unblended Fe II lines for a set of microturbulent
velocities ranging from 0.0 to 2.5 $\kms$. Figure \,\ref{fig1} shows the standard
deviation of the derived [Fe/H] as a function of the microturbulent velocity.
The adopted microturbulent velocity is the value which minimizes the standard deviation ie. for that value, all Fe II lines yield
the same iron abundance.
We therefore adopt a microturbulent velocity $\xi_t$ = 0.96 $\pm$ 0.04 $\kms$ constant with depth  for HR 7098.

\section{The derived abundance pattern for HR 7098}

\begin{figure}[h!]
 \centering
 \includegraphics[width=0.5\textwidth]{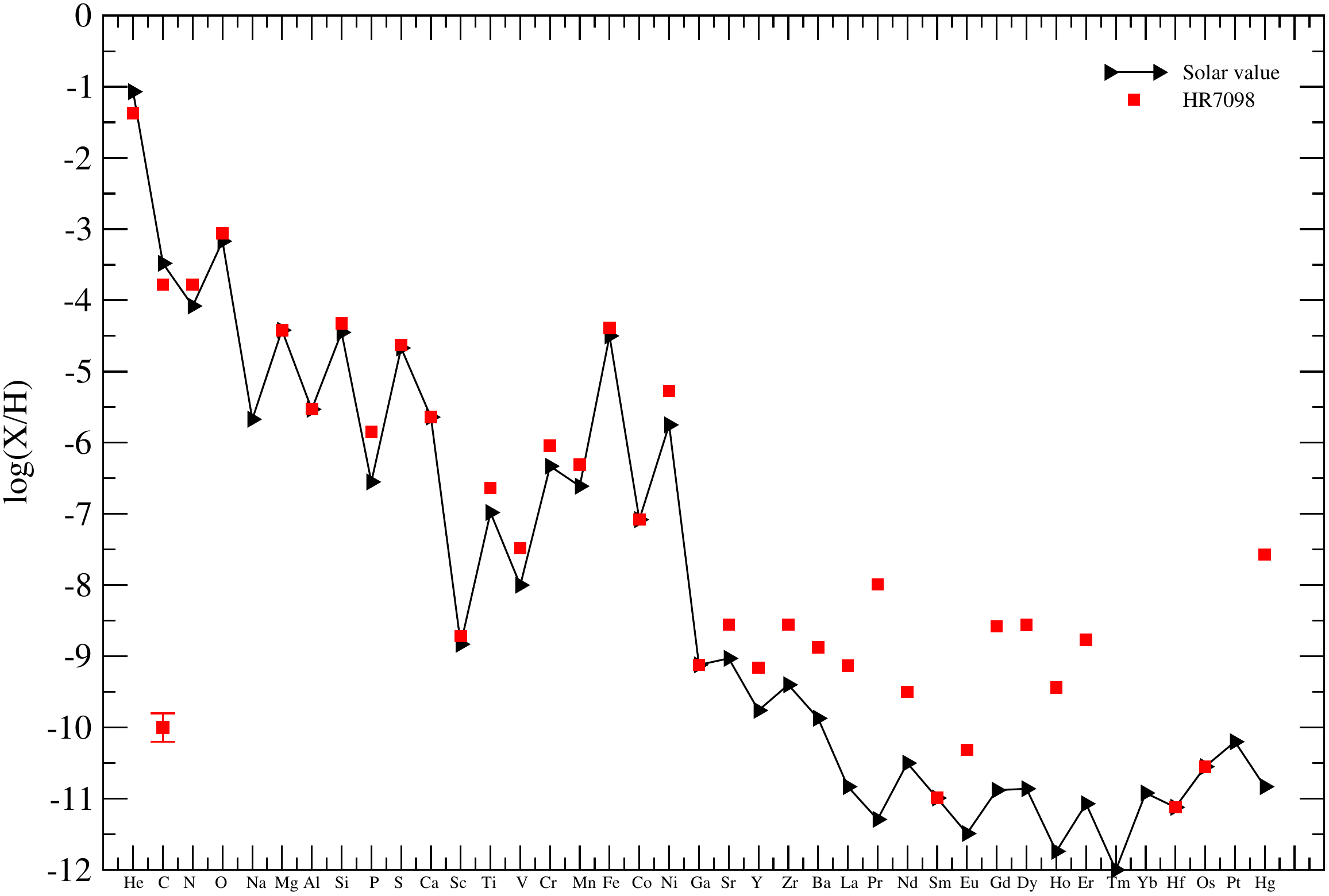}     
  \caption{The derived elemental abundances for HR 7098}
  \label{fig2}
\end{figure}

The derived abundances for the 35 elements studied are displayed in Fig.~\ref{fig2}. For a given element, we display  actually the absolute abundance $\log ({X \over H})$ with a representative error of $\pm$ 0.15 dex. We find that HR 7098 displays underabundances in the light elements He, C. It has solar abundances for Mg, Al, S and Ca and only mild overabundances (less than 5 times the solar values) for P, Ti, V, Cr, Mn, Fe, Ni, Sr, Y and Zr. It has large overabundances (larger than 5 times solar) in several very heavy elements: Ba, La, Pr, Nd, Sm, Eu, Gd, Dy, Ho, Er and Hg. The heaviest element Hg is the most overabundant. The abundance pattern of HR 7098 therefore resembles that of the coolest HgMn stars.


\section{Conclusions}

The derived abundance pattern of HR 7098 departs from the solar composition which definitely shows that HR 7098 is not a superficially normal early A star but is actually another new CP star. We have already reported on the discovery of 5 new CP stars of the HgMn type in \cite{Monier15} and \cite{Monier2016}. HR 7098 has overabundances of both the rare earths and of Hg and its effective temperature and surface gravity place it at the coolest end of the realm of the HgMn stars. Hence we propose that HR 7098 be a new and very cool and mild HgMn star.
We are currently planning more observations of HR 7098 with SOPHIE in order  to complement the abundances derived here and search for putative line variability. This will help us adddress the relationship of HR 7098 to the coolest known HgMn stars and constrain the nature of this interesting new CP star.

\begin{acknowledgements}
The authors acknowledge use of the SOPHIE archive (\url{http://atlas.obs-hp.fr/sophie/}) at Observatoire de Haute Provence. They have used the NIST Atomic Spectra Database and the VALD database operated at Uppsala University (Kupka et al., 2000) to upgrade atomic data.
\end{acknowledgements}



%
\end{document}